\documentstyle[12pt]{article}
\voffset   =  \headheight

\begin{document}

\addtolength{\voffset}{-1in}

\normalsize

\title
{\bf The effective electromagnetic interaction in a dense
fermionic medium in $QED_{2+1}$.}

\author{
{\sf V.V. Skalozub} and {\sf A.Yu. Tishchenko \thanks{E-mail:
tish@phd.dp.ua}}\\
{\sf  Dnepropetrovsk State University, Dnepropetrovk 320625,
 UKRAINE }}

\maketitle

\begin{abstract}
{ The effective Lagrangian of arbitrary varying
in space electromagnetic field in a dense medium is derived. It has been
used for investigation of interaction between charged fermions in the
medium. It is shown the possibility for the formation of metastable electron
bound states in the medium when external magnetic field is applied.}

\end{abstract}

\vskip 0.5cm
{\bf 1 Introduction}

The method of effective Lagrangians has been well developed for
the cases of homogeneous \cite{HE} and/or smoothly varying \cite{Gilkey}
external fields. However, in some problems, for example, such as photon
splitting in the electron-positron plasma \cite{Melr} it is necessary to
consider gauge fields rapidly varying in space or time. For such a type of
conditions the calculation of the effective Lagrangian ($EL$) is not
trivial. In the paper \cite{ST} some important properties and the dependence
on chemical potential $\mu $ of the polarization tensor and the
three-photonic vertex in a dense medium were discovered. The results
obtained allow to construct $EL$ for arbitrary inhomogeneous static
electromagnetic fields. The goal of the present paper is to obtain $EL$ of
electromagnetic field in a dense fermionic medium by making use the
discovered properties of polarization tensors and investigate the
corresponding interaction of charged fermions in a dense environment.
Although proposed procedure has a general character and is expected to be
applicable for various gauge theories, here we shall consider
two-dimensional quantum electrodynamics which is widely used for
investigation of high temperature superconductivity \cite{SC} and quantum
Hall effect \cite{QHE}.

\vskip 0.5cm
{\bf 2 The one-loop effective Lagrangian of electromagnetic
field in a dense fermionic medium}

The effective action of electromagnetic field in the one-loop approximation
can be expressed in terms of infinite series containing polarization tensors
\cite{IZ}
      \begin{eqnarray}\label{ser} {\cal S}^{(n)}(A)&=& \frac{(-1)^n
      e^n}{n}\int\hat{A}(x_1)G(x_2-x_1)\hat{A}(x_2)...
      \hat{A}(x_n)G(x_{1}-x_{n}) d^3 x_1 d^3 x_2 ... d^3 x_n= \nonumber \\
      &~&\frac{(-1)^n e^n}{n(2\pi)^{6n}}
      \int\hat{A}(k_1)G(p_1)...\hat{A}(k_n)G(p_{n})
      e^{ik_1 x_1}e^{ip_1 ( x_2-x_1)}... \times \nonumber \\
      &~&e^{ik_n x_n}e^{ip_n (x_1-x_n)}d^3 x_1... d^3 x_nd^3
      k_1...d^3 k_nd^3 p_1...d^3p_n= \nonumber \\
      &~&\frac{(-1)^n }{n(2\pi)^{3(n-1)}}
      \int A_{\mu_1}(k_1)...{A}_{\mu_n}(k_n)
      \Pi^{\mu_1...\mu_n}(k_1...k_n)\delta(\sum^{n}_{i=1}k_i)
      d^3 k_1... d^3 k_n= \nonumber \\
      &~&\frac{(-1)^n }{n(2\pi)^{3n}}
      \int A_{\mu_1}(k_1)...A_{\mu_n}(k_n)
      \Pi^{\mu_1...\mu_n}(k_1...k_n)
      e^{ix \sum^{n}_{i=1}k_i}d^3 k_1... d^3 k_n d^3 x,
      \end{eqnarray}
where $A_\mu$ is a potential and $G$ is Green's function of electromagnetic
field, $\Pi^{\mu_1...\mu_n}(k_1...k_n)$ are polarization tensors with $n$
external photonic lines carrying momenta $k_i$. For $\Pi^{\mu_1...\mu_n}$
being arbitrary functions of momenta the integration over $k_i$ in eq.(\ref
{ser}) is impossible. However, as it was shown in \cite{ST} for static case
$k_0=0, {\bf k}\neq 0$ in the limit of $\mu \gg m$ the tensors tend to
constants proportional to certain degrees of $\mu$. This important property
allows to integrate over all momenta in eq.(\ref{ser}) and express $EL$ as
follows
\begin{equation}
\label{srl}{\cal L}^{\prime}(A)=\sum^{\infty}_{n=1} \frac{(-1)^n}{n}
A_{\mu_1}(x)...A_{\mu_n}(x)  \Pi^{\mu_1...\mu_n}.
\end{equation}
Moreover, the $\mu$-dependence of tensors occurred to be a decreasing
function of the number of external photonic lines. So, only a few terms with
non-negative degrees of $\mu$ contribute to eq.(\ref{srl})

Now let us consider in the limit $\mu \gg m$ the realization of
eq.(\ref{srl}) for the two-dimensional quantum electrodynamics. In
this case we should take into account in eq.(\ref{srl}) the first three
terms containing tensors $\Pi_{\mu}(k_1), \Pi_{\mu\nu}(k_1,k_2),
\Pi_{\mu\nu\lambda}(k_1,k_2,k_3)$. They have been calculated in \cite{ST}
for static limit for interval of momenta $|{\bf k}| \subset [0; 2\sqrt{
\mu^2-m^2}] $:
\begin{eqnarray}\label{p1}
      \Pi_0=\frac{e}{2\pi}(\mu^2-m^2)\theta(\mu^2-m^2),
      \end{eqnarray}
\begin{eqnarray}
      \bigtriangleup \Pi_{00}
      =-\frac{e^2}{2\pi}
      \theta(\mu^2-m^2)\mu,             \label{pij}
      \bigtriangleup \Pi_{ij}
      =-\frac{e^2}{2\pi}
      \theta(\mu^2-m^2)(\mu-m)
      \left(\delta_{ij}-\frac{k_i k_j}{\mbox{\bf k}^2}\right)
      \end{eqnarray}
\begin{eqnarray}                                 \label{p000}
      \Pi_{000}=\frac{e^3}{\pi}\theta(\mu^2-m^2),
      \Pi_{i0j}=-
      \left(\delta_{ij} -
      \frac{k_i k_j \mbox{\bf k}'^2+k'_i k'_j \mbox{\bf k}^2-
      k_i k'_j(\mbox{\bf k} \mbox{\bf k}')}{\mbox{\bf k}^2 \mbox{\bf k}'^2}
      \right)\frac{\Pi_{000}}{4},
      \end{eqnarray}
where $\bigtriangleup \Pi_{ij}$ is the statistical part of the polarization
tensor which completely determines its properties in the limit considered, $
\theta(x)$ is a step function. In eq.(\ref{p000}) the momentum conservation,
${\bf k}_1+{\bf k}_2+{\bf k}_3=0$, has been taken into account.

As it can be seen tensors have manifestly transversal
structures that have to guarantee the gauge invariance of $EL$.

  Let us consider the electroneutrality condition of the medium
and clarify the question: whether or not the contribution to ${\cal L}
^{\prime}$ of a heavy positively charged subsystem considered as a limiting
form of fermions with  mass $M>m$ cancels the contribution of light
negatively charged system? The answer follows immediately from observation
that all multiphotonic vertices with a number of external lines $n>1$ tend
to zero in the limit $M\to \infty$. The only diagrams giving non-zero
contributions are tadpoles and those contributions can be identified with
usual interaction of static potential with charge. Hence we can have neutral
system with non-zero three photonic interaction.

  Substituting expressions (\ref{p1})-(\ref{p000}) into eq.(\ref
{srl}) one obtains $EL$
\begin{eqnarray}              \label{lag'}
      {\cal L}'(x)&=&B_1A_0(x)-C_1A^{2}_{0}(x)-C_2A_i(x)
      (\delta^{ij}\Delta-\partial^i\partial^j)\tilde{A}_j(x) \nonumber \\
      &+&D_1A_{0}^{3}(x)+D_2A_0(x)
      \left((\delta^{ij}\Delta-\partial^i\partial^j)\tilde{A}_j(x)\right)^2,
      \end{eqnarray}
where notations are introduced: $B=\Pi_0, C_1=\frac{e^2}{4\pi}
\theta(\mu^2-m^2)\mu, C_2=\frac{e^2}{4\pi}\theta(\mu^2-m^2)(\mu-m), D_1=
\frac{e^3}{3\pi}\theta(\mu^2-m^2), D_2=-\frac{e^3}{12\pi}\theta(\mu^2-m^2)$
and $\tilde{A}_j(x)=\frac{1}{(2\pi)^2}\int \frac{A_j(\mbox{\bf k})}{
\mbox{\bf k}^2} e^{i\mbox{\bf k}\mbox{\bf x}}d\mbox{\bf k}$.

 Constructed $EL$ leads to non-linear field equations and can be
used for the arbitrary dependence of $A_\mu(x)$ on spatial coordinates.

\vskip 0.5cm
{\bf 3 Generation of magnetic field in a dense medium}

 Now, as an application of eq.(\ref{lag'}), we are going to
consider the modification of the magnetic field generated by static
electric charge \cite{Kogan} in a dense medium.

 The selfconsistent investigation of this problem requires
consideration of the effective Lagrangian including classical part ${\cal L}
_0=-\frac{1}{4\gamma}F^{2}_{\mu\nu}$, ($\gamma$ is dimensional constant
caused by 2-dimensional nature of theory) and the part generated by
integration over fermionic fields. The latter part contains two kinds of
terms appeared in one-loop approximation. First of them is the vacuum
contribution which includes well-known Chern-Simons term \cite{Jek}, \cite
{Redl} ${\cal L}_{CS}= \frac{m_{CS}}{4\gamma}\epsilon^{\mu\nu\alpha}
F_{\mu\nu}A_\alpha $ ($m_{CS}$-Chern-Simons mass). Second one is the
contribution of medium ${\cal L}^{\prime}$, calculated herein (\ref{lag'}).

 Electric $\phi\equiv A_0$ and vector ${\bf A}$ potentials
generated by the point charge possess axial symmetry $(A_\rho=0, A_\varphi
\neq 0, H(\rho)=\frac{A_\varphi}{\rho}+ \frac{\partial A_\varphi}{\partial
\rho})$ and thus in cylindrical coordinates we have the following equation
of motion
\begin{equation}
      \label{eq11} \Delta \phi+  m_{CS}H(\rho)-2\gamma C_1\phi+  3\gamma
      D_1A^{2}_{0}+  \gamma D_2\bigg(\frac{1}{\rho^2}
      \int^{\rho}_{0}A_\varphi(\rho^{\prime})
      \rho^{\prime}d\rho^{\prime}\bigg)^2=0,
      \end{equation}
\begin{equation}
      \label{eq12}\frac{\partial H}{\partial \rho}+
      m_{CS}\frac{\partial \phi}{
      \partial \rho}-  \gamma C_2\bigg(A_\varphi+ \frac{1}{\rho^2}
      \int^{\rho}_{0}A_\varphi
      (\rho^{\prime})\rho^{\prime}d\rho^{\prime}\bigg)
      -2\gamma D_2\phi \bigg(\frac{1}{\rho^2}
      \int^{\rho}_{0}A_\varphi(\rho^{\prime})
      \rho^{\prime}d\rho^{\prime}\bigg)=0.
      \end{equation}
If we consider the case of large value of the chemical potential and induced
Chern-Simons mass $(\mu\gg m_{CS})$, eqs.(\ref{eq11}),(\ref{eq12}) can be
simplified as follows:
\begin{equation}
      \label{eq21} \Delta \phi-2\gamma C_1\phi=0,
      \end{equation}
\begin{equation}
      \label{eq22}\frac{\partial H}{\partial \rho}+
      m_{CS}\frac{\partial \phi}{
      \partial \rho}- \gamma C_2\bigg(A_\varphi+ \frac{1}{\rho^2}
      \int^{\rho}_{0}A_\varphi(\rho^{\prime})
      \rho^{\prime}d\rho^{\prime}\bigg)=0.
\end{equation}
Actually, this simplification implies that screening of electric potential
in a dense medium is completely determined by chemical potential, and the
Chern-Simons term is important only for generation of magnetic field.

 From eq.(\ref{eq21}) we obtain
      \begin{equation}
      \label{phi} \phi=\gamma e K_0(\lambda\rho),
      \end{equation}
where $K_0$ is MacDonald's function, $\lambda=e\sqrt{\mu\gamma/2\pi}$.

 After substitution eq.(\ref{phi}) in eq.(\ref{eq22}) we have
      \begin{equation} \label{eq_a}
      \frac{\partial^2 A_\varphi}{\partial \rho^2}+
      \frac{1}{\rho}
      \frac{\partial A_\varphi}{\partial \rho}-
      \frac{A_\varphi}{\rho^2}- \gamma
      C_2\bigg(A_\varphi+ \frac{1}{\rho^2}
      \int^{\rho}_{0}A_\varphi(\rho^{\prime})
      \rho^{\prime}d\rho^{\prime}\bigg)=
      m_{CS}\gamma e\lambda K_1(\lambda \rho).
      \end{equation}
By introducing of the new function $h(\rho)=\rho A_\varphi(\rho)$ and
differentiating eq.(12) can be transformed to the linear differential
equation of third order:
      \begin{equation}\label{eq_h}
      \rho h^{\prime\prime\prime}(\rho)-\gamma C_2\rho
      h^{\prime}(\rho)-2\gamma C_2h(\rho)=
      m_{CS}\gamma e\lambda \rho(K_1(\lambda
      \rho)+\lambda\rho K_0(\lambda\rho)).
      \end{equation}
For eq.(\ref{eq_h}) it is possible to derive the asymptotic solution for $
\rho\gg\frac{2}{\sqrt{\gamma C_2}}\sim \frac{4}{e}\sqrt{\frac{\pi}{\gamma\mu}
}$. Obviously, this condition can easily be satisfied as long as we deal
with a high density medium.

 Finally, we obtain the following expression for the magnetic field $
H(\rho)$:
      \begin{eqnarray}\label{expr_h}
      H(\rho)\simeq\pi em_{CS}\gamma e^{-\lambda \rho/\sqrt{2}}
      \left[K_0(\lambda\rho)\mbox{\bf L}_1(\lambda\rho)\Big(\pi+1\Big)+
      K_1(\lambda\rho)\mbox{\bf
      L}_0(\lambda\rho)\Big(\pi+\frac{3}{\lambda\rho}\Big)-
      \frac{6}{\pi}K_1(\lambda\rho) \right]
      \end{eqnarray}
where $\mbox{\bf L}_n$ is modified Struve's function. The leading term in
eq.(\ref{expr_h}) gives the estimation for $H(\lambda\rho)$ at large
distance
\begin{equation}
\label{est_h} H(\rho)\sim  em_{CS}\gamma \frac{e^{-\lambda \rho/\sqrt{2}}}{
\lambda\rho}
\end{equation}

As it is seen, in a dense medium the function describing
generated magnetic field is different from that of at $\mu=0$ \cite{Kogan}
and has a higher rate of decreasing.

\vskip 0.5cm
{\bf 4 Static interaction between the electric
charges in the presence of the external magnetic field}

 At $\mu \neq 0$ the Furry theorem is violated \cite{FR} and
three-photonic interaction mixing $\phi$ and spatial components of potential
${\bf A}$ is not zero (See eq.(\ref{p000})). It should modify the effective
interaction between static charges, especially perceptibly when external
fields is applied.

 For the sake of simplicity let us consider the external
homogeneous magnetic field by potential $A_\varphi=\rho H/2$. As it was
shown above (See eqs.(\ref{expr_h}), (\ref{est_h})), generated magnetic field
decreases very rapidly from the source and can be neglected starting from
the distance $\rho> 1/\lambda \sim \frac{2}{e}\sqrt{\frac{\pi}{\gamma\mu}}$.
In this case we can deal with eq. (\ref{eq11}) only and rewrite it as
follows
      \begin{equation}
      \label{phieq}\Delta \phi+ m_{CS}H-2\gamma
      C_1\phi+\gamma D_2\bigg(\frac{1}{\rho}
      \int^{\rho}_{0}A_\varphi(\rho^{\prime})\rho^{\prime}d\rho^{\prime}
      \bigg)^2=0
      \end{equation}
For this equation one can easily derive the approximate solution, describing
electric potential produced by the point charge:
      \begin{equation}
      \label{fh} \phi(z)\simeq \bigg(1-\frac{\pi m_{CS}}{\gamma e^2}
      \Big(\frac{H}{\gamma e\mu}\Big)z^2+
      \frac{\pi}{72}\Big(\frac{H}{\gamma e\mu}\Big)^2z^4
      \bigg)\gamma  eK_0(z),
      \end{equation}
where $z=\rho/\lambda\simeq \frac{1}{e}\sqrt{\frac{2\pi}{\gamma\mu}}\rho$.
If the Chern-Simons mass is induced one and $\frac{H}{\gamma e\mu}>1$, the
three-photonic interaction term dominates in eq.(\ref{fh}). As it is seen
(Fig.1), potential (\ref{fh}) has a local minimum, which provides an
attraction between electrons at distances dependent on $H$ and $\mu$.

  This picture shows a potential ability of the formation of the
metastable bound electronic states which would be interesting for high
temperature superconductivity.


\vskip 0.5cm
{ \bf 5 Conclusions}

The $QED$ in medium in two spatial dimensions has intensively
been investigated in recent time (See for example \cite{U}). It is expected
that in this way such phenomena as spontaneous magnetization and quantum
Hall effect can be adequately described. In papers \cite{U} the uniform
magnetic field was considered and corresponding effective Lagrangians have
been calculated. The present paper is an endeavour of new approach to the
investigation of more general field configurations which can be considered
in connection with mentioned phenomena.

 The main result of the present paper is the construction of the
gauge invariant $EL$ for static electromagnetic field with an arbitrary
dependence on coordinates. This remarkable possibility can be realized due
to important properties of the polarization tensors \cite{ST}: in a dense
medium the degree of $\mu$ in the asymptotics of the tensors decreases with
an increasing of a number external photonic lines. It provides a rapid
convergence of the series of the one-loop diagrams determining $EL$ owing to
the presence of the small parameter $\sim 1/\mu$. Moreover, just this
property leads to the fact that a few first diagrams in eq.(\ref{ser})
adequately describe an effective non-linear interaction of electromagnetic
field. Obviously, such a dependence has a general character and is not
connected with the number of the spatial dimensions but provided by
structure of the fermionic propagator in medium. Similar procedure can be
realized in other gauge theories: $QED_{3+1}, QCD$. The only necessary
condition for this is the presence of dense environment.

 The most interesting application of proposed $EL$ considered
here is the modification of the electrostatic potential in the presence of
external magnetic field. For some range of $H$ and $\mu$ it is possible to
form metastable electronic bound states in planar structures.

\vskip 1cm
{\bf Acknowledgments}

One of
the authors (A.T.) would like to thank Prof. S.Randjbar-Daemi for reading
the manuscript and the International Centre for Theoretical Physics where
part of this paper has been done for kind hospitality. This work was
partially supported by ISF grant N U1D200.

\vskip 1.5cm

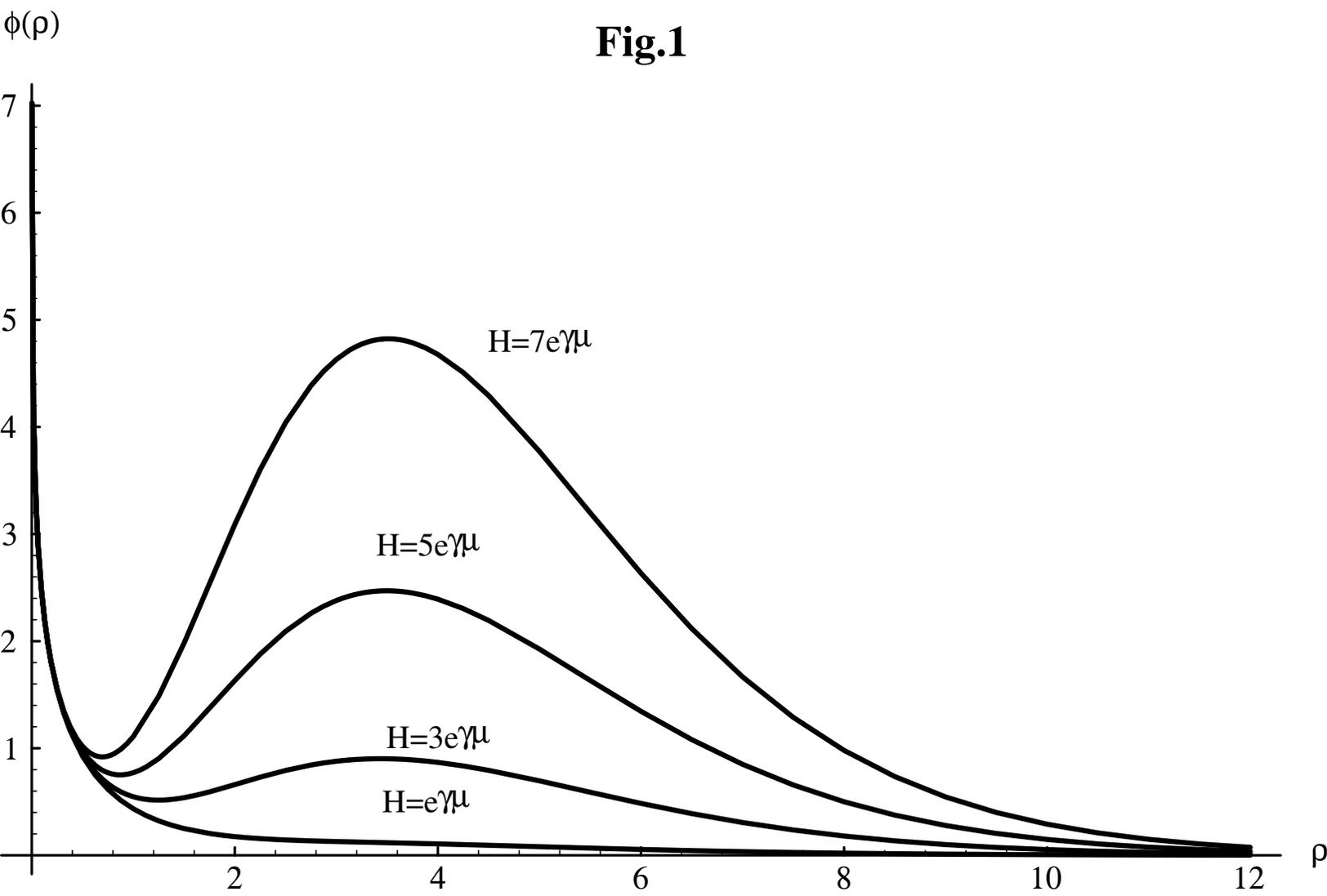
{\large {\bf Figure caption:}\\ (Fig.1.) The electric potential of the point
charge in the presence of the external magnetic field. }


\newpage


\begin{thebibliography}{99}
\bibitem{HE}   W.Heisenberg and H.Euler, {\it Z.Phys.} {\bf 98},
714 (1936)

\bibitem{Gilkey}   J.Schwinger, {\it Phys.Rev.} {\bf 82}, 664
(1951)

\bibitem{Melr}  D.B.Melrose, {\it Plasma Phys.} {\bf 16}, 845
(1974)

\bibitem{ST}  V.V. Skalozub and A.Yu. Tishchenko, {\it JETP}
{\bf 77},  889 (1993)

\bibitem{SC} A.L. Fetter, C.B. Hanna and R.B. Laughlin, {\it %
Phys.Rev.} {\bf B40} 8745 (1989); Y.-H. Chen, F. Wilczek, E. Witten and  B.
Halperin {\it Int.J.Mod.Phys. }{\bf B3} 1001 (1989);  S.Randjbar-Daemi, A.
Salam and J.Strathdee {\it Nucl.Phys.} {\bf B340} 403 (1990); J.E. Hetrick,
Y. Hosotani and  B.-H. Lee {\it Ann.Phys. } {\bf 209} 151 (1991)

\bibitem{QHE}  Eds by R.E. Prangle and S.M.Girvin,{\it Quantum
Hall Effect}  (Springer-Verlag,1987)

\bibitem{IZ} C.Itzykson and J.-B. Zuber, {\it Quantum Field
Theory},  (McGrow-Hill,1980)

\bibitem{Kogan}  Ya.I.Kogan, {\it JETP Lett.} {\bf 49}, 225
(1989)

\bibitem{Jek}  S.Deser, R.Jackiw and S.Templeton {\it Ann.Phys.
}{\bf 140},  372 (1982)

\bibitem{Redl}  A.N. Redlich {\it Phys.Rev.}{\bf D 29} 2326
(1984)

\bibitem{FR}  E.S. Fradkin, {\it Proceedings of Lebedev
Physical  Institute} {\bf 29}, 7 (1965)

\bibitem{U}   D.Weselowski and Y.Hosotani {\it Phys.Lett.}{\bf B 354} 396
(1995);  T.Itoh, SLAC-PUB-DPNU-95-26 (hep-th/9508113); J.O.Andersen and
T.Haugset {\it Phys.Rev.}{\bf D 51} 3073(1995);  S.Kanemura and
T.Matsushita, OU-HET 212 (hep-th/9505146);  D.Cagnemi and E.D'Hoker {\it
Phys.Rev.}{\bf D 51} 2513 (1995);  Vad.Yu.Zeitlin, {\it Phys.Lett.}{\bf B
352} 422 (1995).
\end{thebibliography}
\end{document}